\documentclass[runningheads]{llncs}
\usepackage[T1]{fontenc}
\usepackage{graphicx}
\usepackage{appendix}
\usepackage{bbding}
\usepackage{amsmath}
\usepackage{amssymb}
\newcommand{\A}{\mathcal{A}}
\newcommand{\D}{\mathcal{D}}
\newcommand{\W}{\mathcal{W}}
\usepackage{xcolor}

\usepackage{tikz}
\usetikzlibrary{arrows, quotes}
\tikzstyle{Red Dot}=[fill=red, draw=black, shape=circle]
\tikzstyle{Green Dot}=[fill=white, draw=green, shape=circle]
\tikzstyle{Empty}=[fill=white, draw=black, shape=circle]
\tikzstyle{Black}=[fill=black, draw=black, shape=circle]
\tikzstyle{Red fill}=[-, fill={rgb,255: red,255; green,88; blue,91}]
\tikzstyle{grey fill}=[fill={rgb,255: red,207; green,207; blue,207}, draw=black, shape=circle]
\tikzstyle{Red Circle}=[fill=white, draw=red, shape=circle]
\tikzstyle{letter}=[fill=none, draw=none, shape=circle]

\tikzstyle{Left}=[<-]
\tikzstyle{Right}=[->]

\usetikzlibrary{decorations.pathreplacing}

\pgfdeclarelayer{nodelayer}
\pgfdeclarelayer{edgelayer}
\pgfsetlayers{edgelayer,nodelayer,main}

\usepackage{lineno}

\usepackage{thmtools}
\usepackage{thm-restate}

\begin{document}
%
\title{Tighter Bounds for Wheeler Determinization}
%
%
\author{Philip Bille\inst{1}\orcidID{0000-0002-1120-5154} \and
Inge Li G{\o}rtz\inst{1}\orcidID{0000-0002-8322-4952} \and
Máximo {Pérez-López}\inst{1}\Envelope\orcidID{0009-0001-8507-4820} \and
Simon R. Tarnow\inst{1}\orcidID{0009-0002-4293-6475}}
\authorrunning{P. Bille, I. L. G{\o}rtz, M. {Pérez-López}, S. R. Tarnow}
%
\institute{Technical University of Denmark, 2800 Lyngby, Denmark\\
\email{\{phbi, inge, mpelo\}@dtu.dk}}
\maketitle              
\begin{abstract}
Given a Wheeler NFA $\A$, the \emph{Wheeler determinization} problem is to construct a Wheeler DFA $\D$ that accepts the same language as $\A$. We use the notation $n_\A,m_\A$ for the number of vertices and edges of $\A$, and equivalently $n_\D,m_\D$ for $\D$. Alanko et al. \cite{alanko_regular_2020,alanko_wheeler_2021} solve this problem in $O(n_\A^3)$ time, by constructing a $\D$ that always satisfies $n_\D\leq 2n_\A - 1$. In this paper, we show how to improve the running time to $O(n_\A + m_\A + n_\D + m_\D)$ when the \emph{Wheeler order} of $\A$ is given. If the Wheeler order is not present, we achieve $O(n_\A + m_\A\log n_\A + n_\D + m_\D)$ time by using an algorithm of Becker et al. \cite{becker_sorting_2023}.

Our running time is a factor $n_\A^2/\sigma$ faster than the state of the art for sorted inputs, where $\sigma$ is the size of the alphabet. Furthermore, for $\sigma=O(1)$ we have the first linear time algorithm for this problem. We show that our bound is tight with any combination of $n_\A$ and $\sigma$, by giving a family of inputs for which our output $\D$ is minimum, and of maximum size $\Theta(n_\A\sigma)$.

\end{abstract}
%
%
%



\section{Introduction}
Wheeler graphs \cite{gagie_wheeler_2017} are a class of directed, labelled graphs that has received attention in both the compression community and bioinformatics. In compression, Wheeler graphs allow storing a (potentially infinite) set of strings, and support indexing for matching patterns $P$ in $O(|P|\log \sigma)$ time, in contrast to the conditional lower bound of $\omega(m^\delta|P|^\beta)$ time (for $\delta<1$ or $\beta< 1$) required in general labelled graphs \cite{equi_graphs_2023}. They have also been studied as an important subclass of finite automata \cite{alanko_regular_2020,alanko_wheeler_2021}. In the bioinformatics community, Wheeler graphs have been used for indexing genomic databases. The compressed index of the variation graph toolkit VG \cite{garrison_variation_2018} is based on the Wheeler properties of the de Bruijn graph of the data, as well as other indices such as VARI \cite{muggli_succinct_2017} and Themisto \cite{maklin_bacterial_2021}.

Most of the algorithmic work on Wheeler graphs has been described on \emph{deterministic} Wheeler graphs, that is, graphs where no two edges with the same label leave the same vertex. Important examples are computing matching statistics \cite{conte_computing_2023}, computing the LCP array \cite{alanko_computing_2024}, counting distinct $k$-mers \cite{alanko_computing_2026}, and minimizing the size \cite{alanko_linear-time_2022}. Further work that extends results from Wheeler graphs to arbitrary finite automata also maintains the deterministic restriction \cite{kim_faster_2023,cotumaccio_prefix_2023}, or provides efficient algorithms only for deterministic automata \cite{cotumaccio_indexing_2021}. It is therefore an important question whether these results can be extended to non-deterministic graphs, and under what conditions. 

In what follows, we use the notation $n_\W$ and $m_\W$ for the number of vertices and edges of an automaton $\W$, respectively, and define $|\W|=n_\W+m_\W$.
Alanko et al. describe, in their seminal papers \cite{alanko_regular_2020,alanko_wheeler_2021}, a deterministic Wheeler automaton (WDFA) with the same language of any non-deterministic Wheeler automaton (WNFA), dubbed the \emph{Wheeler determinization} of the WNFA. It has at most $2n_\A-1-\sigma$ nodes and $O(n_\A\sigma)$ edges, where $\sigma$ is the size of the alphabet. They give an $O(n_\A^3)$ time algorithm to construct the Wheeler determinization, which is not output-sensitive given the stated size. Still, this result is remarkable, given that in general graphs the blow-up in size when going from NFAs to DFAs is exponential in the worst case.

Their algorithm works on WNFAs without needing their Wheeler order. We focused on improving the running time of the algorithm using the properties of Wheeler graphs, and making use of the Wheeler order. If the input is not sorted, we can use the algorithm of Becker et al. \cite{becker_sorting_2023} to compute first an equivalent, sorted version of $\A$ with the same language. 


\paragraph*{Our contribution.}We describe an algorithm for building the WDFA described by Alanko et al. that runs in output-sensitive time when the Wheeler order is given. More precisely:
\begin{theorem}\label{thm:determinization}
    Given an input WNFA $\A$, there is an algorithm to construct its Wheeler determinization $\D$ in $O(n_\A+m_\A\log n_\A+|\D|)$ time. If the Wheeler order is given, we can construct $\D$ in $O(|\A| + |\D|)$ time.
\end{theorem}
Focusing on sorted inputs, given that $m_\W=O(n_\W\sigma)$ holds for any Wheeler graph \cite{gibney_complexity_2022}, we have a linear time algorithm for constant alphabets. If all vertices of $\A$ are reachable from the source (a standard assumption, see \cite{alanko_regular_2020}), then the number of edges of $\D$ dominates the running time, and we obtain an improvement of $n_\A^2/\sigma$ over the state of the art.
Additionally, we show that the upper bound is tight for any $n_\A$ and $\sigma$, by showing a family of WNFAs where our determinization $\D$ is the minimum WDFA with the same language as $\A$, and its size is $\Theta(n_\A\sigma)$.

\begin{corollary}\label{corollary:tight}
    There exist inputs for the Wheeler determinization problem that require worst-case time $\Omega(n_\A\sigma)$. There is an algorithm for the problem that runs in $O(n_\A\sigma+m_\A\log n_\A)$ time, or in optimal $\Theta(n_\A\sigma)$ time if the Wheeler order is given.
\end{corollary}

Our approach follows the classical powerset construction algorithm to build a DFA with the same language as an input NFA. To achieve our results, we introduce two new data structures: firstly, an edge discovery data structure that outputs the distinct edge labels outgoing from an interval of nodes; and secondly, a bitvector-based dynamic set data structure for intervals of nodes.


\section{Preliminaries}
We work with directed, labelled graphs $G=(V, E)$, where the edges have single character labels. The labels come from a finite alphabet $\Sigma$ of size $\sigma$ with a total order $\preceq$. The edges are a set $E\subseteq V\times V\times \Sigma$, and we denote them by a triple $(u,v,a)$ where $u$ is the \emph{tail} and $v$ is the \emph{head}. We assume standard notions of nondeterministic and deterministic finite state automata; see, e.g. \cite{hopcroft2001introduction}. We denote a finite automaton by $\A=(V,E,\Sigma,s,F)$ and its language by $\mathcal{L}(\A)$. We use the notation $|V|=n_\A,|E|=m_\A$ and $|\A|=n_\A+m_\A$ throughout the paper, and we drop the subscript when it is understood from the context.
The set $\textrm{Pref}(\mathcal{L}(\A))$ is the set of all prefixes of strings in $\mathcal{L}(\A)$.

\paragraph*{Wheeler graphs \cite{gagie_wheeler_2017}.}A Wheeler graph $\A=(V,E)$ is a directed, labelled graph with a total order $\leq$ on the vertices, called the Wheeler order, that satisfies the following properties. For any two edges $(u, v,a),(u',v',a')\in E$:
\begin{description}
    \item[W1:]\label{W1} $a\prec a'\implies v\leq v'$
    \item[W2:] $(a=a')\wedge (u<u')\implies (v\leq v')$
\end{description}
Additionally, sources come before vertices with positive in-degree in the Wheeler order. Wheeler graphs have the property of \emph{path coherence} \cite{gagie_wheeler_2017}: given an interval of vertices $[i,j]\subseteq V$ and any string $\alpha$, the subset of vertices that we reach when following walks that spell $\alpha$ starting from $[i,j]$ is also an interval of $V$. Wheeler graphs also have the property that $m_\A=O(n_\A\sigma)$ (see \cite[Theorem 4]{gibney_complexity_2022}).

A Wheeler NFA (WNFA) is an NFA whose state graph is a Wheeler graph. A Wheeler DFA (WDFA) is a WNFA with a deterministic state graph. Given a WNFA $\A$, the \emph{Wheeler determinization} of $\A$ is a WDFA $\D = (V^\D,E^\D,\Sigma,s^\D,F^\D)$ with $\mathcal{L}(\D)=\mathcal{L}(\A)$, defined as follows  \cite{alanko_regular_2020,alanko_wheeler_2021}. Let $I_\alpha\subseteq V$ be the interval of vertices that we reach when following walks that spell a string $\alpha$, starting from $s$. Then, $V^\D=\{I_\alpha \mid \alpha\in \textrm{Pref}(\mathcal{L}(\A))\}$, $E^\D=\{(I_\alpha, I_{\alpha e},e)\in V^\D\times V^\D\times \Sigma\}$, $s^\D=I_{\epsilon}=[s,s]$, and $F^\D=\{I_\alpha:\alpha\in \mathcal{L}(\A)\}$. 
Note that this WDFA may not be the WDFA with the minimum size for the language $\mathcal{L}(\A)$. However, we have a bound on the number of vertices: $n_\D\leq 2n_\A-1-\sigma$  (see \cite{alanko_regular_2020,alanko_wheeler_2021}).

A Wheeler graph with $n$ vertices can be represented by a succinct data structure (called \emph{index}) for navigation and pattern matching queries. We describe the two components that we use. Let $\ell_i=\deg_{out}(v_i)$ be the out degree of the $i$-th vertex in Wheeler order, and let $L_i$ be the sequence of its outgoing labels, ordered by $\preceq$. We define the bitvector $O=0^{\ell_1}10^{\ell_2}1\cdots0^{l_n}1$, and a character vector $L=L_1L_2\cdots L_n$. 
\paragraph*{Rank space.}
We can assume that the labels of any WNFA $\A$ are in rank space (meaning $\Sigma\subseteq [1,m]$) if $\A$ is sorted, even if the alphabet size is unbounded. To show this, represent the edges of $\A$ by pairs $(v, u)\rightarrow a$, where $v$ is the head and $u$ is the tail of the edge, and the label $a$ is only satellite data. Now, radix sort the pairs, and do a linear scan to re-label the satellite data in $[1,m]$ following the new found order. By property W1, this corresponds to the natural order of the labels.
Because the Wheeler order gives indices in $[1,n]$ to the vertices, the radix sort step takes linear time.
\paragraph*{Sorting assumption. }If the input WNFA $\A$ is not sorted, we can use the algorithm of Becker et al. \cite{becker_sorting_2023} to find a total pre-order on the vertices in $O(m_\A\log n_\A)$ time. Constructing the quotient automaton with respect to the equivalence relation induced by the pre-order produces a WNFA $\A'$ with $\mathcal{L}(A')=\mathcal{L}(\A)$ that we can then use to determinize. The pre-order is equivalent to a total, Wheeler order on $\A'$. Thus, for our purposes, we can assume the input is sorted at a $O(m_\A\log n_\A)$ cost.
\paragraph*{Data structures.}
We use standard Range Minimum Query (RMQ) data structures of $O(n)$ space, constructible in $O(n)$ time and with $O(1)$-time queries \cite{fischer_optimal_2010,bender_lca_2000}. We also use succinct rank and select data structures for bit-vectors, that use $n+o(1)$ space, are constructible in $O(n)$ time and have $O(1)$ query time \cite{JacobsonPhDRank,ClarkSelect}.
\section{Data structures}
In this section we describe two data structures specialized for a Wheeler NFA~$\A$.
The first is a dynamic set data structure for the intervals of $\A$'s vertices, with worst-case deterministic constant time for search and insertion.
The second is an edge discovery data structure that answers the following query: given an interval of vertices $[v_i,v_j]\subseteq V$, let $\Lambda_{ij}$ be the set of labels on the outgoing edges from $[v_i,v_j]$. For each $a\in \Lambda_{ij}$, it returns the maximal intervals of vertices that we reach by traversing the edges labelled $a$ starting from $[v_i,v_j]$. 
This last data structure is based on the colored range query data structure of Muthukrishnan \cite{muthukrishnan_efficient_2002}, which we also review here for completeness.

\subsection{Dynamic set data structure for intervals of vertices}\label{subsec:dynamic_set}
Given a WNFA $\A$, in \cite{alanko_regular_2020,alanko_wheeler_2021} it was proven that the vertices of the determinization $\D$ are elements of a \emph{prefix/suffix family of intervals} $\mathcal{C}$ over $[1,n]$. The elements are intervals $I=[m_I,M_I]\subseteq [1,n]$, where $m_I$ is the smallest endpoint, and $M_I$ is the largest. We use the following order of the intervals: $I<J$ iff $(m_I<m_J)\vee (m_I=m_J\wedge M_I<M_J)$. If we consider the sum of the endpoints for any two intervals $I<J$ in this special family $\mathcal{C}$, it holds that $m_I+M_I<m_J+M_J$. This implies that there are at most $2n-1$ elements in $\mathcal{C}$, and that we have a total order for them given by $rank(I)=m_I+M_I$. 

Since we allow ourselves to use $O(n)$ space, and we want to represent a dynamic set of size at most $O(n)$, we can use a bit vector $B[1,2n-1]$ as our set data structure. To add $I$ to $B$, set $B[m_I+M_I]$ to $1$. An interval $J$ of the family is in $B$ if and only if $B[m_J+M_J]$ is set to $1$. We can also support deletions if needed, by setting $B[m_I+M_I]$ to $0$. Since $I\neq J$ implies $m_I+M_I\neq m_J+M_J$, the queries are correct. Summarizing:

\begin{lemma}\label{lemma:dynamic_set}
    If $\mathcal{C}$ is a prefix/suffix family of intervals over $[1,n]$, there exists a dynamic set data structure that implements search, insertion and deletion over elements of $\mathcal{C}$ in $O(1)$ worst-case time and $2n-1$ bits of space.
\end{lemma}

\subsection{Colored range queries}\label{subsection:colored_range_queries}
We review the \emph{colored range} data structure of Muthukrishnan \cite{muthukrishnan_efficient_2002}. Given an array $A[1,n]$ of elements from a universe of size $u=O(n)$, we preprocess the array to solve the following query: given a range of indices $[i,j]\subseteq [1,n]$, return the set of unique elements in $A[i,j]$. In practice, our data structure will report the indices of such elements in $A$, that we can retrieve later.

To preprocess the array, we build a \emph{predecessor array} $P[1,n]$ that holds, in the $i$-th element, the index of the previous occurrence of $A[i]$ in $A$, or $-1$ if it is the first occurrence. Then, we build a \emph{Range Minimum Query} (\emph{RMQ}) data structure \cite{fischer_optimal_2010,bender_lca_2000} over $P$. 
The predecessor array can be built by scanning $A$ and keeping track of the last occurrence of each character of the alphabet with an array of size $u$, in time $O(n+u)=O(n)$. 
The RMQ data structure can also be built in linear time \cite{fischer_optimal_2010,bender_lca_2000}.

To solve the colored range query over $[i,j]$, we query the index $w$ of the minimum element in $P[i,j]$ with the RMQ data structure. If $P[w]<i$, then the previous occurrence of $A[w]$ in $A$ is outside the range $[i,j]$, therefore we report $w$. To continue, we recurse on $[i,w-1]$ and $[w+1,j]$, and we keep on reporting the index of elements as long as their previous occurrence appears before $i$. This reports all the indices of the $occ$ unique elements in $A[i,j]$ in $O(1+occ)$ time. 

We make a crucial observation: the indices that are found by the colored range query are the indices of the \emph{first} occurrence of elements in $A[i,j]$, because any further occurrence of an element in $A[i,j]$ has a predecessor index in $[i,j]$, and thus it will not be reported.

\begin{restatable}[Muthukrishnan, 2002]{theorem}{muthukrishnan}\label{thm:muthukrishnan}
    Given an array $A[1,n]$ over a universe of size $O(n)$, we can preprocess $A$ in $O(n)$ time to build a linear space data structure that answers colored range queries in $O(1+occ)$ time. The indices reported are the first occurrences of each element in the queried interval. 
\end{restatable}

\subsection{Edge discovery data structure}\label{subsec:edge_discovery}
We augment the data structure of Muthukrishnan to create a data structure for edge discovery queries. We need one last building block: observe that we can obtain easily a colored range query that reports the indices of the \emph{last} occurrences of the set of characters inside a range. To achieve this, in the construction of the data structure, instead of a predecessor array we use a \emph{successor} array $S[1,n]$, where $S[i]$ is the index of the next occurrence of $A[i]$ in $A$, or $n+1$ otherwise. Then, instead of range minimum queries, we use range \emph{maximum} queries to obtain the maximum indices of the distinct characters in a range. The query for a range $[i,j]$ is similar: as long as the retrieved index has a successor larger than $j$, we recurse both left and right. 

To support edge discovery queries, at preprocessing time, build the two versions of colored range queries on $L$. We also augment $L$ by adding to each label the vertex at the head of the edge. This extra information can be added in linear time, by sorting the edges in the format $(u,v,a)$ with radix sort and extracting the heads (the resulting order is the same as the order in $L$). Then, build an array $E[1,\sigma]$ to hold pairs of vertices (initially empty), and create an empty queue of vertices $Q$. We also build the standard rank and select data structures \cite{JacobsonPhDRank,ClarkSelect} on the bitvector $O$.

To solve an edge discovery query for an interval $[v_i,v_j]$, we first obtain the indices in $L$ of the first outgoing edge from $v_i$ and the last outgoing edge from $v_j$, denoted $[i_e,j_e]$, by $i_e=rank_0(select_1(i-1,O),O)+1$, $j_e=rank_0(select_1(j,O),O)$. 
Then, we do a colored range query for $[i_e,j_e]$ on $L$, to find the indices of the first occurrences in $L[i_e,j_e]$ of the set of unique elements in $L[i_e,j_e]$. For each reported index $w$, we add $a=L[w]$ to the queue $Q$. We also retrieve the vertex at the head of the edge labelled by $L[w]$, and we set it as the first element of the pair at $E[a]$. Then, we repeat the query on $L[i_e,j_e]$ but using the other colored range data structure that finds the last occurrences inside the range. For each reported index $w'$, we retrieve the vertex at the head of the edge labelled by $L[w']$, and we add it to the second element of the pair at $E[L[w']]$. Finally, we pop one by one the characters in $Q$, and for each one $a$, we report $(a, E[a])$.

It follows from Theorem \ref{thm:muthukrishnan} and the rank space assumption that we can build this data structure in $O(|\A|)$ time, and that an edge discovery query will run in $O(1+occ)$ time, where $occ$ is the number of pairs $(a,[v_i',v_j'])$ that we output. Note that, since we only inspect the indices of $E$ that we output after we have set them to the correct values, we can support a future edge discovery query without any further processing or cleanup.

\begin{lemma}\label{lemma:edge_discovery}
    Given a sorted WNFA $\A$, we can build an edge discovery data structure in $O(|\A|)$ time and space that answers queries in $O(1+occ)$ time.
\end{lemma}

\section{Determinization algorithm}
In this section we describe the $O(|\A|+|\D|)$ algorithm to compute the Wheeler determinization $\D$ given the sorted WNFA $\A$. 

The input to our algorithm will be the bit-vector $O$ and the character vector $L$ as described before. These can be built in linear time once we know the Wheeler order of $\A$. We can also assume a bit-vector $B_F[1,n_\A]$ that is $1$ if the $i$-th node of $\A$ is final, and $0$ otherwise. The data structures we need are as follows. We first make a queue $K$ to hold newly found vertex intervals, and we add $[v_1,v_1]$ at the start. We also build the dynamic set data structure $H$ (see section \ref{subsec:dynamic_set}) for intervals of vertices, and insert $[v_1,v_1]$ in it. Then, we build the edge discovery data structure $\mathcal{E}$ that we described in section \ref{subsec:edge_discovery}. Finally, we build a constant-time rank data structure for $B_F$. The output will be computed as the adjacency list representation of the output WDFA graph. Neither the vertices of $\D$ nor their adjacency lists are guaranteed to be sorted at the end, but once the graph is deterministic, it can be sorted in linear time if needed \cite[Theorem 3.3]{alanko_regular_2020}.

The algorithm runs until the queue $K$ is empty. The main loop pops an interval $[v_i,v_j]$ from $K$, and discovers the set of outgoing intervals of vertices $[v_i',v_j']$ with each label $a\in \Sigma$, using a query to $\mathcal{E}$. For each outgoing interval $[v_i',v_j']$, we first check in the set $H$ if $[v_i',v_j']$ exists already. If so, we just add the edge $([v_i,v_j],[v_i',v_j'],a)$ to the adjacency list of $[v_i,v_j]$. Otherwise, we add that edge, add $[v_i',v_j']$ to $H$, and enqueue $[v_i',v_j']$ in $K$. When all the outgoing edges from $[v_i,v_j]$ are processed, we return to the start of the loop.

Now we can construct the set of final states easily. We determine if $[v_i,v_j]\in V^\D$ is final by computing $rank_1(v_j,B_F)-rank_1(v_{i-1},B_F)$ (we set $rank_1(v_0,B_F)=0$). If the result is greater than 0, it means that there exists at least one final state in $[v_i,v_j]$, and thus we set $[v_i,v_j]$ as a final state; otherwise we set it as non-final. 

\subsubsection{Analysis.}
The correctness follows from the same argument as the powerset construction algorithm, see for example \cite[Lemma 2.6]{alanko_regular_2020}. 
We need to argue the time complexity. Building $H$ takes $O(n_\A)$ time by Lemma \ref{lemma:dynamic_set}, and building $\mathcal{E}$ takes $O(|\A|)$ time by Lemma \ref{lemma:edge_discovery}. The bit-vector $B_F$ is built in $O(n_\A)$ time. This amounts to $O(|\A|)$ time.
During the construction algorithm we do $n_\D$ iterations of the main loop. Each iteration with a vertex $[v_i,v_j]$ makes a query to $\mathcal{E}$ that takes $O(1+\deg_{out}([v_i,v_j]))$ time, and for each reported output, we do at most one lookup and one insertion into $H$, plus at most one insertion into $K$. All the operations on $H$ and $K$ are constant time. Thus, the time to run all iterations adds up to $O(n_\D+m_\D)=O(|\D|)$ time. Finally, identifying final states takes $O(n_\D)$ time. In total, we arrive at $O(|\A|+|\D|)$ time. If the Wheeler order is not given, prior to determinization we use the algorithm by Becker et al. \cite{becker_sorting_2023} and run our algorithm on the output WNFA, at a $O(m_\A\log n_\A)$ cost. This proves Theorem \ref{thm:determinization}.

Since $n_\D\leq 2n_\A-1-\sigma$, $m_\D=O(n_\D\sigma)$ and $m_\A=O(n_\A\sigma)$, we have $|\D|=O(n_\A\sigma)$ and $O(|\A|+|\D|)=O(n_\A\sigma)$. For constant alphabets, this implies that our running time is $O(n_\A)$. This proves the upper bound of Corollary \ref{corollary:tight}.
\section{Tight family}



In this section we show a family of WNFAs parameterized by $\sigma$, such that for any WNFA $\A$ in the family, the output WDFA $\D$ of our algorithm is minimum, and has $n_\D\leq 2n_\A - 1 - \sigma$ vertices and $m_\D=\Theta(n_\A\sigma)$ edges. Since Wheeler graphs always satisfy $m=O(n\sigma)$, this output is as large as possible. Figure \ref{fig:wnfa} shows an example WNFA with $\sigma=O(n_\A)$, and Figure \ref{fig:wdfa} shows the output $\D$. We can construct instances for an arbitrary $\sigma$ by including only $\sigma-2$ different nodes in the bottom row, each with a unique incoming label; and we increase $n$ by making the top row longer. It is easy to see that $\D$ will always have $O(n)$ edges from the top row to each of the $\sigma-2$ nodes in the bottom row. We consider all nodes final.

Minimality of $\D$ (see \cite{alanko_regular_2020,alanko_linear-time_2022} for the precise conditions) can be seen in two steps: firstly, the top row vertices are inequivalent because they have a distinct outgoing string ending at $v_{10}$. Secondly, the bottom row vertices are inequivalent due to their distinct incoming labels. This proves the lower bound of Corollary \ref{corollary:tight}.

\begin{figure}[hbt]
    \centering
    \begin{minipage}{0.45\columnwidth}
\begin{tikzpicture}
	\begin{pgfonlayer}{nodelayer}
		\node [style=Empty] (0) at (-1.5, 0.75) {a};
		\node [style=Empty] (1) at (-0.5, 0.75) {a};
		\node [style=Empty] (2) at (0.5, 0.75) {a};
		\node [style=Empty] (3) at (1.5, 0.75) {a};
		\node [style=Empty] (5) at (-1.5, -0.25) {b};
		\node [style=Empty] (6) at (-0.5, -0.25) {c};
		\node [style=Empty] (7) at (0.5, -0.25) {d};
		\node [style=Empty] (8) at (1.5, -0.25) {e};
		\node [style=letter] (9) at (-2.25, 1.25) {$v_1$};
		\node [style=letter] (10) at (-1.5, 1.25) {$v_2$};
		\node [style=letter] (11) at (-0.5, 1.25) {$v_3$};
		\node [style=letter] (12) at (0.5, 1.25) {$v_4$};
		\node [style=letter] (13) at (1.5, 1.25) {$v_5$};
		\node [style=letter] (14) at (-1.5, -0.75) {$v_6$};
		\node [style=letter] (15) at (-0.5, -0.75) {$v_7$};
		\node [style=letter] (16) at (0.5, -0.75) {$v_8$};
		\node [style=Empty] (17) at (-2.25, 0.75) {$\epsilon$};
		\node [style=letter] (18) at (1.5, -0.75) {$v_9$};
		\node [style=Empty] (19) at (2.5, 0.75) {f};
		\node [style=letter] (20) at (2.5, 1.25) {$v_{10}$};
	\end{pgfonlayer}
	\begin{pgfonlayer}{edgelayer}
		\draw [style=Left, in=135, out=45, loop] (0) to ();
		\draw [style=Right] (0) to (1);
		\draw [style=Right] (1) to (2);
		\draw [style=Right] (2) to (3);
		\draw [style=Right] (0) to (5);
		\draw [style=Right] (0) to (6);
		\draw [style=Right] (0) to (7);
		\draw [style=Right] (0) to (8);
		\draw [style=Right] (17) to (0);
		\draw [style=Right] (3) to (19);
	\end{pgfonlayer}
\end{tikzpicture}
    \caption{Example Wheeler NFA $\A$}
    \label{fig:wnfa}
    \end{minipage}
    \begin{minipage}{0.45\columnwidth}
\begin{tikzpicture}
	\begin{pgfonlayer}{nodelayer}
		\node [style=Empty] (0) at (-1.5, 0.75) {a};
		\node [style=Empty] (1) at (-0.5, 0.75) {a};
		\node [style=Empty] (2) at (0.5, 0.75) {a};
		\node [style=Empty] (3) at (1.5, 0.75) {a};
		\node [style=Empty] (5) at (-1.5, -0.25) {b};
		\node [style=Empty] (6) at (-0.5, -0.25) {c};
		\node [style=Empty] (7) at (0.5, -0.25) {d};
		\node [style=Empty] (8) at (1.5, -0.25) {e};
		\node [style=letter] (9) at (-2.5, 1.25) {$[v_1, v_1]$};
		\node [style=letter] (10) at (-1.5, 1.25) {$[v_2, v_2]$};
		\node [style=letter] (11) at (-0.5, 1.25) {$[v_2,v_3]$};
		\node [style=letter] (12) at (0.5, 1.25) {$[v_2,v_4]$};
		\node [style=letter] (14) at (-1.5, -0.75) {$[v_6.v_6]$};
		\node [style=letter] (15) at (-0.5, -0.75) {$[v_7,v_7]$};
		\node [style=letter] (16) at (0.5, -0.75) {$[v_8.v_8]$};
		\node [style=Empty] (17) at (-2.5, 0.75) {$\epsilon$};
		\node [style=letter] (18) at (1.5, 1.25) {$[v_2,v_5]$};
		\node [style=letter] (19) at (1.5, -0.75) {$[v_9,v_9]$};
		\node [style=Empty] (20) at (2.5, 0.75) {f};
		\node [style=letter] (21) at (2.5, 1.25) {$[v_{10},v_{10}]$};
	\end{pgfonlayer}
	\begin{pgfonlayer}{edgelayer}
		\draw [style=Right] (0) to (1);
		\draw [style=Right] (1) to (2);
		\draw [style=Right] (2) to (3);
		\draw [style=Right] (0) to (5);
		\draw [style=Right] (1) to (6);
		\draw [style=Right] (2) to (7);
		\draw [style=Right] (3) to (8);
		\draw [style=Right] (1) to (5);
		\draw [style=Right] (2) to (5);
		\draw [style=Right] (3) to (5);
		\draw [style=Right] (2) to (6);
		\draw [style=Right] (3) to (6);
		\draw [style=Right] (3) to (7);
		\draw [style=Right, in=-30, out=-75, loop] (3) to ();
		\draw [style=Right] (17) to (0);
		\draw [style=Right] (0) to (6);
		\draw [style=Right] (0) to (7);
		\draw [style=Right] (0) to (8);
		\draw [style=Right] (1) to (7);
		\draw [style=Right] (1) to (8);
		\draw [style=Right] (3) to (20);
		\draw [style=Right] (2) to (8);
	\end{pgfonlayer}
\end{tikzpicture}
    \vspace*{-10mm}
    \caption{Its Wheeler determinization $\D$}
    \label{fig:wdfa}
    \end{minipage}
\end{figure}


 \begin{credits}
\subsubsection{\ackname} This work was also supported by Independent Research Fund Denmark grants 10.46540/3105-00302B and 10.46540/4283-00129B. We thank the anonymous reviewers for a thorough analysis and useful comments. The third author also thanks Jarno N. Alanko for his introduction to the topic of this paper.
 \end{credits}

%
%
%

\bibliographystyle{splncs04}
\bibliography{WheelerFiltered}





\end{document}